\documentclass[aps,preprintnumbers,eqsecnum,amsmath,amssymb,showpacs,nofootinbib]{revtex4}
\usepackage{graphicx}
\usepackage{dcolumn}
\usepackage{bm}
\usepackage{epsfig}

\begin{document} 
\title{The puzzle of the {\boldmath $\pi\to\gamma\,\gamma^*$} transition form factor}
\author{Wolfgang Lucha$^{1}$ and Dmitri Melikhov$^{1,2,3}$}
\affiliation{
$^1$HEPHY, Austrian Academy of Sciences, Nikolsdorfergasse 18, A-1050 Vienna, Austria\\
$^2$Faculty of Physics, University of Vienna, Boltzmanngasse 5, A-1090 Vienna, Austria\\
$^3$SINP, Moscow State University, 119991 Moscow, Russia}
\date{\today}
\begin{abstract}
We study the $P\to\gamma\,\gamma^*$ ($P=\pi^0,\eta,\eta'$)
transition form factors by means of the local-duality (LD) version
of QCD sum rules. For the case of $\eta$ and $\eta'$, the conventional LD model  
provides a good description of~the existing data. However, for the
$\pi$ form factor we find disagreement with recent {\sc BaBar}~results 
for high $Q^2$ even though the accuracy of the LD approximation is expected to increase with $Q^2$. 
It remains mysterious why the $\eta$ and $\eta'$ form factors to virtual photons, on the
one hand, and the $\pi$ form~factor, on the other hand, show a
qualitatively different behaviour corresponding to a rising with $Q^2$ violation of 
local duality in the pion case. In a quantum mechanical example we show that, 
for a bound-state size of~about 1 fm, the LD sum rule provides an accurate prediction 
for the form factor for $Q^2\ge$ a few~GeV$^2$.
\end{abstract}
\maketitle


\section{Introduction}
\begin{figure}[hb!]
\begin{center}
\begin{tabular}{c}
\includegraphics[width=8.4cm]{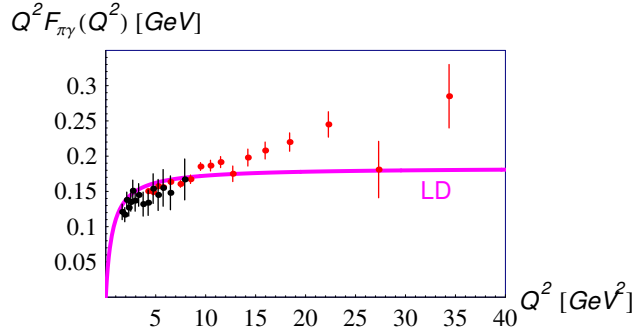}
\end{tabular}
\caption{\label{Plot:0}
Form factor $F_{\pi\gamma}$ as function of $Q^2$: experimental data from 
\cite{babar} (red) and 
\cite{cello,cleo} (black); 
the solid line represents the interpolation~(\ref{bl}).}
\end{center}
\end{figure}

The form factor describing the two-photon transition of a light pseudoscalar meson $P$ is one of the 
simplest hadronic form factors in QCD. The corresponding amplitude
\begin{eqnarray}
\langle\gamma(q_1)\gamma(q_2)|P(p)\rangle&=&i
\epsilon_{\varepsilon_1\varepsilon_2 q_1 q_2}
F_{P\to\gamma\gamma}(q_1^2,q_2^2)
\end{eqnarray}
contains only one invariant form factor, $F_{P\to\gamma\gamma}(q_1^2,q_2^2)$. 
We shall be interested in the case of one real and one virtual photons, 
$q_1^2=0$ and $-q_2^2=Q^2\ge0,$ and define
$F_{P\gamma}(Q^2)\equiv F_{P\to\gamma\gamma}(q_1^2=0,q_2^2=-Q^2).$
For the pion case, the form~factor $F_{\pi\gamma}(Q^2)$ has the
following properties: 
(i) In the chiral limit of massless quarks
and a massless pseudoscalar $\pi$, the form factor at $Q^2=0$ is
given by the axial anomaly \cite{abj}, 
$F_{\pi\gamma}(Q^2=0)=1/(2\sqrt{2}\pi^2f_\pi)$, $f_\pi=130$ MeV. (ii) At large $Q^2$,
perturbative QCD (pQCD) predicts the asymptotic behaviour
\cite{bl} $F_{\pi\gamma}(Q^2\to\infty)\to \sqrt{2}f_\pi/Q^2$. 
Brodsky and Lepage proposed a simple interpolating formula between
these two values \cite{bl}
\begin{eqnarray}
\label{bl}
F_{\pi\gamma}(Q^2)=\frac{1}{2\sqrt{2}\pi^2f_\pi}
\left(1+\frac{Q^2}{4\pi^2f_\pi^2}\right)^{\!-1}.
\end{eqnarray}
This formula may not provide an accurate description of the form
factor at small nonzero $Q$ but one may expect its accuracy to
increase rather fast at $Q^2$ values larger than a few GeV$^2$. In
contrast to these expectations, the~{\sc BaBar} collaboration
recently reported a surprising result for the behaviour of the
$F_{\pi\gamma}$ form factor \cite{babar}: the product $Q^2
F_{\pi\gamma}(Q^2)$ does not saturate at large $Q^2$ but increases
further (Fig.~\ref{Plot:0}).

This result has already attracted a lot of attention in the literature (see, e.g.~\cite{kroll} and 
references therein). 
The aim of our analysis is to study the $P\gamma$ transition form
factor for $P=\pi^0,\eta,\eta'$ by making use of the local-duality
version of QCD sum rules \cite{ld}. This approach allows one to study hadron form factors without knowing subtle 
details of their structure and to consider on equal footing form factors of different hadrons. 

We find that for both $\eta$
and $\eta'$ transitions, the LD sum rule provides a satisfactory description of the form factors in the region
$Q^2=5-100$ GeV$^2$ \cite{cello,cleo,babar2,babar1}. The result from the
LD approximation for the $\pi$~case, on the other hand, does not agree with the {\sc BaBar} data.

In order to test the accuracy of the LD sum rule for the $P\gamma$
transition form factor, we consider, in parallel~to~QCD, a
quantum-mechanical potential model. In the latter case, the form
factor can be obtained both by the LD sum rule and by an exact
calculation. Comparing these results with each other provides a
probe of the LD approximation \cite{blm}. Here we find
that, independently of the details of the confining potential, the
LD sum rule reaches the accuracy~of a few percent
already at relatively low values of the momentum transfer.

It remains mysterious why, in contrast to the success of the LD
approximation for the $\eta$ and $\eta'$ transitions and~to~the
experience from quantum mechanics, the results from the LD approximation 
strongly contradict the {\sc BaBar} result.

This Letter is organized as follows: 
In the next section, we recall the structure of the $\langle VAV \rangle$
amplitude~and the relation of $F_{\pi\gamma}$ to the axial anomaly. 
Section 3 presents the dispersion representation for $\langle VAV \rangle$ and touches the 
issue of multiloop radiative corrections to this quantity. 
Section 4 reviews the LD approximation for the $P\gamma$ form factor and 
studies the expected accuracy of this approximation making use of a quantum-mechanical testing ground. 
Section 5 applies the LD sum rule to the $F_{P\gamma}$, $P=\pi^0,\eta,\eta'$ form factors. 
Section 6 summarizes our conclusions.


\section{The three-point function {\boldmath $\langle VAV\rangle$} and the axial anomaly}
Let us start with the amplitude of two-photon production from the
vacuum induced by the axial-vector current~of nearly massless
quarks of one flavour, $j_\mu^5=\bar q\gamma_\mu\gamma_5q$, 
with $\varepsilon_{1,2}$ denoting the photon polarization vectors:
\begin{equation}
\langle\gamma(q_1)\gamma(q_2)|j_\mu^5(x=0)|0\rangle=
T_{\mu\alpha\beta}(p|q_1,q_2)\varepsilon^\alpha_1\varepsilon^\beta_2,
\qquad p=q_1+q_2.
\end{equation}
The amplitude $T_{\mu\alpha\beta}$
is obtained from the vacuum expectation value of the $T$-product
of two vector and one axial-vector currents and will be referred to
as the $\langle VAV\rangle$ amplitude. Vector-current conservation
yields the following relations:
\begin{eqnarray}
T_{\mu\alpha\beta}(p|q_1,q_2)q_1^\alpha=0,\qquad
T_{\mu\alpha\beta}(p|q_1,q_2)q_2^\beta=0.
\end{eqnarray}
The general decomposition of the amplitude contains four invariant form factors and may be written as
\begin{eqnarray}
\label{amp1}
T_{\mu\alpha\beta}(p|q_1,q_2)&=&
-\frac{p_\mu}{p^2-m_P^2}\epsilon_{\alpha\beta q_1q_2}iF_A
+(q_1^2\epsilon_{\mu\alpha\beta q_2}-q_{1\alpha}\epsilon_{\mu q_1\beta q_2}-\frac{p_\mu}{p^2-m_P^2}q_1^2\epsilon_{q_1\alpha \beta q_2})iF_1
\nonumber\\
&&+(q_2^2\epsilon_{\mu\beta\alpha q_1}-q_{2\beta}\epsilon_{\mu q_2\alpha q_1}-\frac{p_\mu}{p^2-m_P^2}q_2^2\epsilon_{q_2 \beta \alpha q_1})iF_2
+(q_1-q_2)_\mu\epsilon_{\alpha\beta q_1 q_2}iF_3.
\end{eqnarray}
The explicit calculation of these form factors at one
and two loops in QCD leads to $F_3=0$ \cite{2loop}. Therefore, we
shall~omit the corresponding term. The absence of any contact
terms in $T_{\mu\alpha\beta}$ can be verified by reducing out one
of the photons and using the conservation of the electromagnetic
current \cite{ms}.

The parameterization (\ref{amp1}) takes into account the pole at
$p^2=m_P^2$, related to the contribution of the lightest pseudoscalar state. 
Apart from the pole at $p^2=m_P^2$, the amplitude has no singularities at small $p^2$. In the chiral
limit, $m_P=0$, both the second and third Lorentz structures in (\ref{amp1}) are transverse with respect to $p_\mu$, 
and the
form~factor $F_A$ represents the axial anomaly \cite{abj}.
According to the Adler--Bardeen theorem \cite{ab}, the axial
anomaly is saturated~by~the one-loop expression, $F_A=1/(2\pi^2)$,
and remains non-renormalized by higher-order corrections.

Separating the longitudinal and the transverse structures for the case $m_P\ne 0$, we obtain
\begin{eqnarray}
\label{amp2}
T_{\mu\alpha\beta}(p|q_1,q_2)&=&
-\frac{p_\mu}{p^2}\epsilon_{\alpha\beta q_1 q_2}i F_A  - \frac{p_\mu}{p^2(p^2-m_P^2)}m_P^2\epsilon_{\alpha\beta q_1 q_2}i (F_A+q_1^2F_1+q_2^2F_2) 
\nonumber \\
&&+(q_1^2\epsilon_{\mu\alpha\beta q_2}-q_{1\alpha}\epsilon_{\mu q_1\beta q_2}-\frac{p_\mu}{p^2}q_1^2
\epsilon_{q_1 \alpha \beta q_2})i F_1 
\nonumber \\ 
&&+(q_2^2\epsilon_{\mu\beta\alpha q_1}-q_{2\beta} \epsilon_{\mu q_2\alpha
q_1}-\frac{p_\mu}{p^2}q_2^2\epsilon_{q_2 \beta \alpha q_1})i F_2.
\end{eqnarray}
Beyond the chiral limit, the first two terms contain
two poles \cite{lmope}: a ``kinematical'' pole at $p^2=0$, which
cancels the corresponding singularities in the transverse Lorentz
structures, and the ``dynamical'' pole at $p^2=m_P^2$
corresponding to the $\pi$ meson. The full amplitude is regular at
$p^2=0$. That is, the pole at $p^2=0$ in the transverse
Lorentz~structures is of purely kinematic origin and does not
correspond to a massless particle. By forming the divergence, one
obtains
\begin{eqnarray}
\label{div}
ip^\mu T_{\mu\alpha\beta}=\epsilon_{\alpha\beta q_1q_2}
\left[F_A-\frac{m_P^2}{m_P^2-p^2}(F_A+q_1^2F_1+q_2^2F_2)\right].
\end{eqnarray}
As is clear from (\ref{div}), the transition form factor of
interest is given by the linear combination of $F_A,$ $F_1,$ and $F_2$
\begin{eqnarray}
\label{2.6}
F_{P\to\gamma\gamma}=-\frac{1}{f_{P}}(F_A+q_1^2F_1+q_2^2F_2)|_{p^2=m_P^2}.
\end{eqnarray}
Thus, the $P\to\gamma\,\gamma$ form factor is proportional to the
axial anomaly only at one kinematical point, $q_1^2=q_2^2=0$. At
this~point, the pion form factor in the chiral limit is protected
from radiative corrections by the Adler--Bardeen theorem; the one-loop result represents the exact result. 
Hence,~in the chiral limit one expects the pion pole at $p^2=0$ to emerge in one-loop diagrams for 
the $\langle VAV\rangle$ amplitude. Indeed, in this unique situation a pole dual to a
single hadron state emerges from the single one-loop diagram of perturbation theory \cite{zakharov}.

However, already if one of the photons is virtual, the transition form
factor $F_{P\to\gamma\gamma}$ and the axial anomaly are not proportional to each other. 
It is the topic of the next two sections to analyze
the form factor $F_{P\gamma}(Q^2)$ by means of dispersive sum rules~\cite{svz}.


\section{Dispersion representations for {\boldmath $\langle VAV\rangle$} and the axial anomaly}
We now discuss the one-loop expression for the amplitude. 
To this end, a slightly different parameterization, obtained by setting 
$(F_A+q_1^2F_1+q_2^2F_2)/(p^2-m_P^2)=F_0,$ proves to be convenient:
\begin{eqnarray}
T_{\mu\alpha\beta}(p|q_1,q_2)=-p_\mu \epsilon_{\alpha\beta q_1 q_2}i F_0 
+(q_1^2\epsilon_{\mu\alpha\beta q_2}-q_{1\alpha}\epsilon_{\mu q_1\beta q_2})i F_1
+(q_2^2\epsilon_{\mu\beta\alpha q_1}-q_{2\beta} \epsilon_{\mu q_2\alpha q_1})i F_2.
\end{eqnarray}
As follows from (\ref{2.6}), $F_0$ contains the contribution of the pseudoscalar meson of our interest. 

We also consider the transition amplitude of the pseudoscalar current operator $\bar q\gamma_5q$:
\begin{eqnarray}\label{G5}
\langle\gamma(q_1)\gamma(q_2)|\bar q\gamma_5q|0\rangle=
\epsilon_{\alpha\beta q_1q_2}\varepsilon^\alpha_1\varepsilon^\beta_2 F_5(q_1^2,q_2^2,p^2).
\end{eqnarray}
The two-photon amplitude of the divergence of the axial current takes the form
\begin{eqnarray}
\label{anomaly1}
\langle\gamma(q_1)\gamma(q_2)|\partial^\mu j^5_\mu|0\rangle=
\epsilon_{\alpha\beta q_1 q_2}\varepsilon^\alpha_1 
\varepsilon^\beta_2 (p^2 F_0-q_1^2 F_1-q_2^2F_2).
\end{eqnarray}
The case of our interest is $q_1^2=0$, then the form factor $F_1$
does not contribute to the divergence. In perturbation theory, the
form factors $F_0$, $F_2$, and $F_5$ may be written in terms of
their spectral representations in $p^2$ (with $q^2\equiv q_2^2=-Q^2$):
\begin{eqnarray}
F_i(p^2,q^2)=\frac{1}{\pi}
\int\limits_{4m^2}^\infty\frac{ds}{s-p^2}\,\Delta_i(s,q^2).
\end{eqnarray}
To one-loop order, the spectral densities read \cite{ms,m,teryaev1,teryaev2}
\begin{eqnarray}
\label{1loop}
\Delta_0(s,q^2)&=&-\frac{1}{2\pi}\frac{1}{(s-q^2)^2}
\left[-q^2\,w+2m^2\log\left(\frac{1+w}{1-w}\right)\right],\nonumber\\
\Delta_2(s,q^2)&=&-\frac{1}{2\pi}\frac{1}{(s-q^2)^2}
\left[-s\,w+2m^2\log\left(\frac{1+w}{1-w}\right)\right],\nonumber\\
\Delta_5(s,q^2)&=&-\frac{1}{2\pi}\frac{m}{s-q^2}
\log\left(\frac{1+w}{1-w}\right),\qquad w\equiv\sqrt{1-4m^2/s}.
\end{eqnarray}
Obviously, the absorptive parts $\Delta_i$ obey the classical
equation of motion for the divergence of the axial current
\begin{eqnarray}
s\,\Delta_0(s,q^2)-q^2\,\Delta_2(s,q^2)=2m\,\Delta_5(s,q^2).
\end{eqnarray}
The form factors then satisfy
\begin{eqnarray}
\label{3.7}
p^2F_0(p^2,q^2)-q^2F_2(p^2,q^2)=2m\,F_5(p^2,q^2)
-\frac{1}{\pi}\int\limits_{4m^2}^\infty ds \,\Delta_0(s,q^2).
\end{eqnarray}
The last integral is equal to $-{1}/{2\pi}$, independently of the values of
$m$ and $q^2$, and represents the axial anomaly \cite{abj}:
\begin{eqnarray}
\label{dd}
p^2F_0(p^2,q^2)-q^2F_2(p^2,q^2)=2m\,F_5(p^2,q^2)+\frac{1}{2\pi^2}.
\end{eqnarray}
In the chiral limit $m=0$ and for $q^2=0$, the form factor $F_0$
develops a pole related to a massless pseudoscalar~meson \cite{zakharov}. 
The residue of this pole is again the axial-anomaly $1/{2\pi^2}$. 

As is clear from (\ref{3.7}), the anomaly represents the integral of $\Delta_0$, 
the spectral density of the form factor $F_0$. Adler and Bardeen tell us that the anomaly is 
non-renormalized by multiloop corrections. The easiest realization of this property would have been just 
the vanishing of multiloop contributions to the spectral density $\Delta_0(s,q^2)$. An argument in favour of 
this possibility comes from explicit two-loop calculations \cite{teryaev1,2loop} which report the  
non-renormalizability of the full $\langle VAV\rangle$ vertex to the two-loop accuracy. 
However, if so, the full form factor $F_0$ is given by its one-loop expression. Then, this expression should 
develop the pion pole, known to be present in the full amplitude for any value of $q^2$. 
But, obviously, this pole does not emerge in the one-loop expression for $F_0$ if $q^2\ne 0$! 

This requires that multiloop corrections to the form factor $F_0$ (and, respectively, to its absorptive part
$\Delta_0(s,q^2)$) do not vanish. Then one may ask oneself how it may happen that the anomaly nevertheless remains 
non-renormalized by multiloop corrections? The only possible answer we see is that the non-renormalization 
of the anomaly is reached due to some conspiral property of multiloop contributions to $\Delta_0(s,q^2)$ 
forcing their integral to vanish 
(in fact, quite similar to the one-loop result: although the spectral density explicitly depends on $m$ and $q^2$, 
its integral is an $m$- and $q^2$-independent constant). 

So we conclude that,~in spite~of the fact that explicit calculations yield a
non-renormalizability of the full $\langle VAV\rangle$ vertex to two-loop accuracy \cite{teryaev1,2loop}, 
the conjecture of \cite{2loop} that this result might hold to all orders of the perturbative expansion may not be valid.


\section{ {\boldmath $F_{P\gamma}$ from a local-duality sum rule for $F_0$}}
As is obvious from (\ref{amp2}), the contribution of the light pseudoscalar 
constitutes a part of the form factor $F_0$. 
The Borel sum rule for the corresponding Lorentz structure reads ($Q^2=-q^2>0$)
\begin{eqnarray}
\int ds\exp(-s\tau)\Delta_0(s,Q^2)=-f_P F_{P\gamma}(Q^2)\exp(-m_P^2\tau)+\mbox{contributions of excited states}.
\end{eqnarray}
Exploiting the concept of duality, the contribution of the excited
states is assumed to be dual to the high-energy region of the
diagrams of perturbation theory above an effective threshold
$s_{\rm eff}$. After that, setting the Borel parameter $\tau=0$
(which yields the so-called local-duality limit), we arrive at the
LD sum rule for a pseudoscalar $\bar q q$-meson 
\begin{eqnarray}
\label{ld1}
\int\limits_{4m^2}^{s_{\rm eff}(Q^2)}ds\,\Delta_0(s,Q^2)=- f_P F_{P\gamma}(Q^2).
\end{eqnarray}
The spectral density $\Delta_0$ to one-loop order is given by (\ref{1loop}); two-loop corrections were 
found to be absent \cite{teryaev1,2loop}. As discussed above, higher-loop corrections to $\Delta_0$ 
cannot vanish; so the l.h.s. of 
(\ref{ld1}) is known to $O(\alpha_s^2)$ accuracy. 
Recall that all details of the nonperturbative dynamics are encoded
in a single quantity, the effective threshold $s_{\rm eff}(Q^2)$.
The effective threshold is an essential parameter of the method of dispersive sum rules; it is not identical to 
the physical threshold and is not universal (i.e., it is specific for the correlator under consideration); 
moreover, in general it depends on $Q^2$ \cite{lms}.

In the chiral limit, the LD expression for the form factor for the one-flavour case is particularly simple:
\begin{eqnarray}
\label{ld}
F_{P\gamma}(Q^2)=\frac{1}{2\pi^2f_P}\frac{s_{\rm eff}(Q^2)}{s_{\rm eff}(Q^2)+Q^2}.
\end{eqnarray}
Apart from neglecting $\alpha_s^2$ and higher-order corrections to the spectral density $\Delta_0$, 
no approximations have been done up to now: we have just considered the LD limit $\tau=0$; 
for an appropriate choice of $s_{\rm eff}(Q^2)$ the form factor may still be calculated exactly. 
Approximations come into the game when we consider a model for $s_{\rm eff}(Q^2)$. 

Irrespective of the behaviour of $s_{\rm eff}(Q^2)$, at $Q^2=0$ the form factor is related to the axial anomaly: 
$F_{P\gamma}(0)=1/(2\pi^2f_P)$. QCD factorization requires $s_{\rm eff}(Q^2)\to 4\pi^2f_P^2$ for large $Q^2$. 
The simplest model compatible with this requirement is obtained by setting
\begin{eqnarray}
\label{ldappr}
s_{\rm eff}(Q^2)= 4\pi^2f_P^2
\end{eqnarray}
for {\it all} values of $Q^2$ \cite{rad2}. We shall refer to the choice (\ref{ldappr}) yielding for the neutral 
pion case the Brodsky--Lepage result (\ref{bl}), as the ``conventional LD model''.\footnote{In an alternative 
approach to the $P\gamma$ form factor \cite{mikhailov},  
the pseudoscalar meson is described by a set of distribution amplitudes of increasing twist which are 
treated as nonperturbative inputs. 
In our analysis, the deviation of the effective threshold $s_{\rm eff}(Q^2)$ from its asymptotic value $4\pi^2 f_P^2$ 
corresponds to some extent to the contribution of higher-twist distribution amplitudes in the 
approach of \cite{mikhailov}.}
The relevance and the expected accuracy of the LD model may be tested in those cases where the form factor 
$F_{P\gamma}(Q^2)$ is known, i.e., may be calculated by other theoretical approaches or measured
experimentally. Then, the exact effective~threshold may be reconstructed from (\ref{ld}), 
in this way probing the accuracy of the LD model. Let us play this game making use of a
quantum-mechanical model, where the exact form factor may be calculated from the solution of 
the Schr\"odinger equation, and then compared with the result of the sum rule for a three-point
function of nonrelativistic field theory.

\subsection{LD sum rule in quantum mechanics}
The analogue of the
$\pi\gamma$ form factor in quantum mechanics is given by
\cite{qmsr,lms2}
\begin{eqnarray}
\label{fnr}
F_{\rm NR}(\bm{q})=\int\limits_0^\infty dT
\langle\Psi|J(\bm{q})\exp(-HT)|\bm{r}=\bm{0}\rangle,
\end{eqnarray}
where the current operator $J(\bm{q})$ is introduced by the kernel
$\langle\bm{r}'|J(\bm{q})|\bm{r}\rangle=
\exp(i\bm{q}\cdot\bm{r})\,\delta^{(3)}(\bm{r}-\bm{r}'),$ the
Hamiltonian $H$ governing the nonrelativistic potential model
reads
\begin{eqnarray}
H=\frac{\bm{k}^2}{2m}+V(r),
\end{eqnarray}
and $\Psi$ is the corresponding ground state. Technically, the LD
model for the form factor (\ref{fnr}) is constructed from the quantum-mechanical analogue 
of the three-point function in the same way as for the case of the elastic form factor (for details, consult \cite{blm}). 
Recall, however, an essential conceptual difference between the $P\gamma$ form factor and the
elastic form factor with respect to the factorization of these
quantities at large momentum transfers: The factorization of the
elastic form factor requires the presence of both Coulomb and confining terms in the interaction. 
The factorization of the $P\gamma$ form factor does not require the presence of a Coulombic term and 
emerges also for a purely confining interaction. The LD model for a given form factor is tightly
related to its factorization properties; specifically, the LD
model for the analogue of the $P\gamma$ form factor may be formulated in quantum
mechanics for the case of a purely confining~potential.

\begin{figure}[b]
\begin{center}
\begin{tabular}{c}
\includegraphics[width=7.7cm]{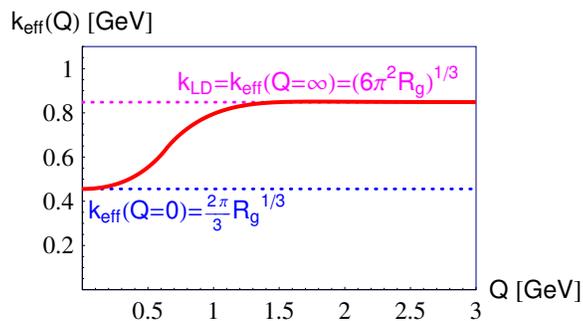}
\end{tabular}
\caption{\label{Plot:1}
Effective threshold $k_{\rm eff}(Q)$ for the three-point function in quantum mechanics, recalculated 
from the exact form factor $F_{\rm NR}(Q)$, Eq.~(\ref{fnr}), in the harmonic-oscillator potential model. 
$R_g\equiv |\Psi(r=0)|^2$.}
\end{center}
\end{figure}

Figure \ref{Plot:1} depicts the exact effective threshold $k_{\rm eff}(Q)$ ---  
the quantum-mechanical counterpart~of the effective threshold $s_{\rm eff}(Q^2)$ --- for the example 
of the harmonic-oscillator potential $V(r)=m\omega^2 r^2/2,$ for parameter values relevant for hadron 
physics: a reduced mass of the light quark of $m=0.175$ GeV and an interaction strength~of $\omega=0.5$ GeV, 
which lead to a size of the ground state around $1$ fm, a typical size of a ground-state hadron in QCD. 
We have checked that a similar picture for the effective threshold emerges for other confining potentials; 
moreover, adding the Coulomb potential changes this picture only slightly.

From the behaviour of $k_{\rm eff}(Q),$ we conclude that the LD model may be expected to work increasingly 
well already~for $Q^2$ above a few GeV$^2$. Inspired by this result, we now look what the LD model 
predicts for $(\pi,\eta,\eta')\to\gamma\,\gamma^*$ transitions.


\section{\boldmath The $(\pi^0,\eta,\eta')\to\gamma\,\gamma^*$ form factor}

\subsection{\boldmath The $\pi^0\to\gamma\,\gamma^*$ form factor}
\begin{figure}[ht!]
\begin{center}
\begin{tabular}{c}
\includegraphics[width=8.4cm]{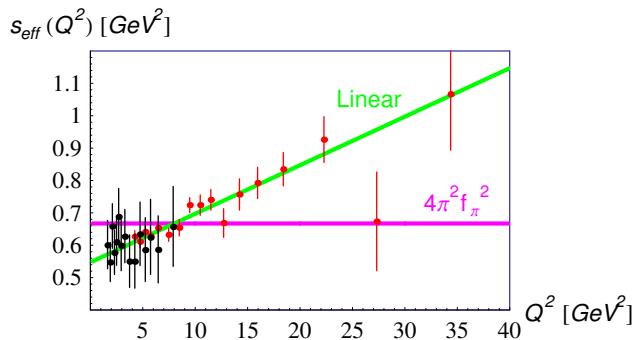}\end{tabular}
\caption{\label{Plot:2}Effective threshold $s_{\rm eff}(Q^2)$
recalculated from the data \cite{babar} (red) and \cite{cello,cleo} (black)  
by means of the LD relation (\ref{Fpi}) for the form factor~$F_{\pi\gamma}(Q^2)$.}
\end{center}
\end{figure}

Taking into account the $\pi^0$ flavour structure and choosing the relevant interpolating current 
$j_\mu^5=(\bar u\gamma_\mu\gamma_5u-\bar d\gamma_\mu\gamma_5d)/\sqrt{2}$, the LD sum rule reads  
\begin{eqnarray}
\label{Fpi}
F_{\pi^0\gamma}(Q^2)=
\frac{N_c}{\sqrt{2}}\left(\frac49-\frac19\right)\frac{1}{2\pi^2f_P}\frac{s_{\rm eff}(Q^2)}{s_{\rm eff}(Q^2)+Q^2}, \qquad N_c=3,  
\end{eqnarray}
which by setting $s_{\rm eff}(Q^2)=4\pi^2 f_\pi^2$, $f_\pi=130$ MeV, leads to the Brodsky-Lepage formula (\ref{bl}). 
Figure~\ref{Plot:0} shows the corresponding plot. 
Figure \ref{Plot:2} represents~the ``experimental'' effective
threshold recalculated from the form factor data via
Eq.~(\ref{Fpi}). This ``experimental'' effective threshold may be
well approximated by a linearly rising function of $Q^2$.
Surprisingly, the {\sc BaBar} data show very~strong --- and
growing with $Q^2$ --- violations of local duality! This
observation is in absolute contradiction to our experience from
quantum mechanics. Let us investigate next what happens in the
case of the $\eta$ and $\eta'$ mesons.

\vspace{-0.5cm}

\subsection{\boldmath The $(\eta,\eta')\to\gamma\,\gamma^*$ form factor}
\begin{figure}[ht!]
\begin{center}
\begin{tabular}{cc}
\includegraphics[width=8.6cm]{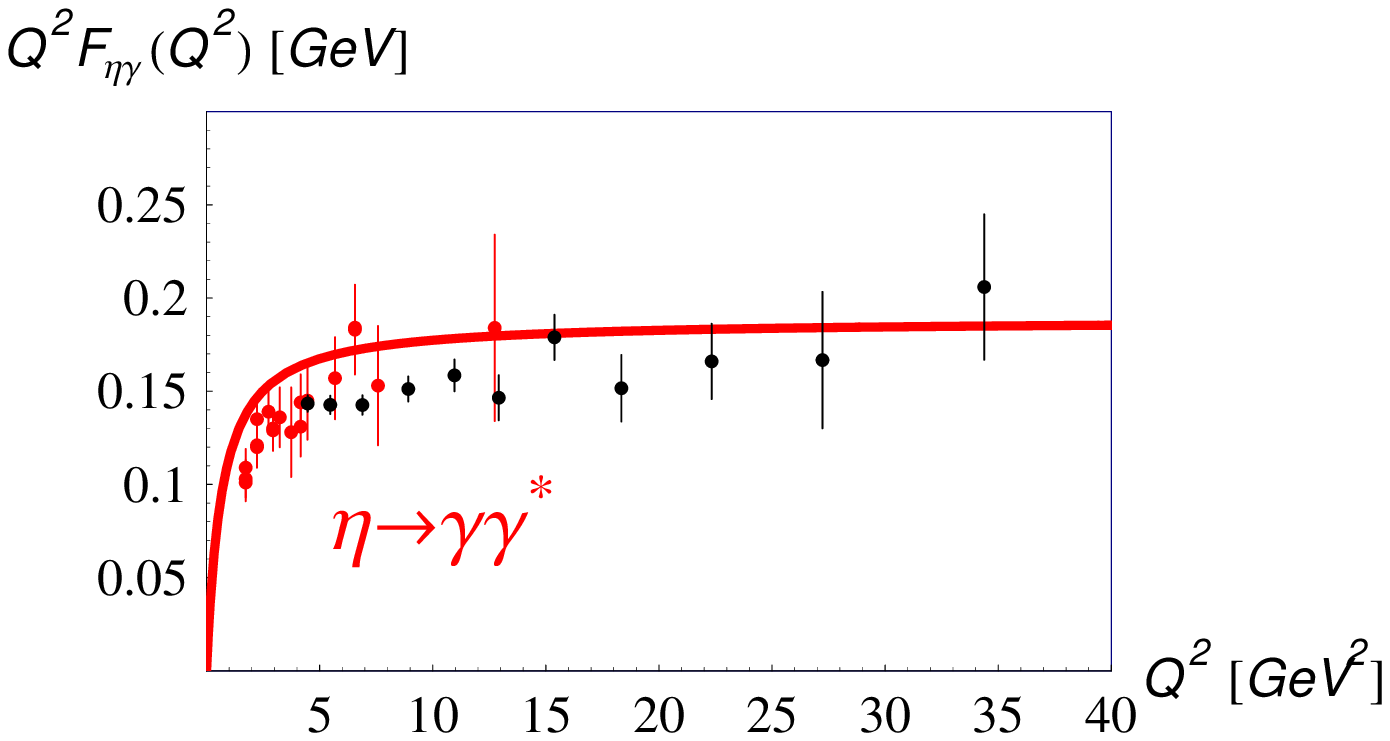}&
\includegraphics[width=8.6cm]{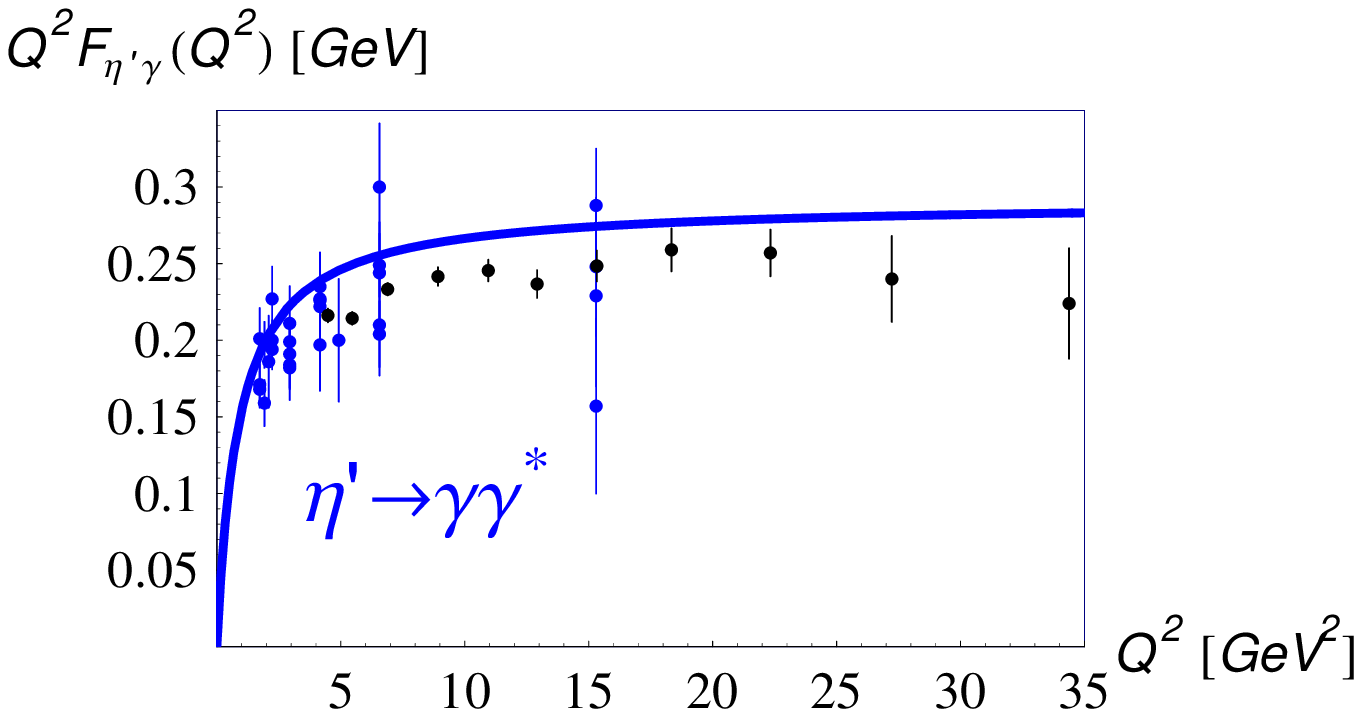}\\
\end{tabular}\caption{\label{Plot:3} LD form factors
$F_{\eta\gamma}$ and $F_{\eta'\gamma}$ vs.\ $Q^2$, compared with the experimental data presented in
\cite{cello,cleo} (coloured dots) and \cite{babar1} (black dots). 
The data points ``borrowed'' from the timelike momentum transfer $q^2=-Q^2=112$ GeV$^2$ \cite{babar2}, 
$q^2F_{\eta\gamma}(q^2)=0.229\pm 0.03\pm 0.008$ GeV and 
$q^2F_{\eta'\gamma}(q^2)=0.251\pm 0.019\pm 0.008$ GeV are not shown in this plot.}
\end{center}
\end{figure}

The simple expression (\ref{bl}) is sometimes erroneously assumed also
for the $\eta$ and $\eta'$ cases \cite{dorokhov,cleo}. However,
the~na\"ive replacement $f_\pi\to f_{\eta,\eta'}$ in (\ref{bl})
yields wrong expressions for $F_{(\eta,\eta')\gamma}$. The correct
way to proceed is to take into~account the presence of two ---
nonstrange and strange --- components in the $\eta$ and $\eta'$
mesons and their mixing. Making use~of the $\eta$--$\eta'$ mixing
scheme from \cite{anisovich,feldmann} (see also \cite{chernyak}),
the flavour structure of $\eta$ and $\eta'$ may be 
described as follows
\begin{eqnarray}
|\eta\rangle &=&|\frac{\bar uu+\bar dd}{\sqrt{2}} \rangle\cos \phi-|\bar ss\rangle \sin \phi,\nonumber\\
|\eta'\rangle&=&|\frac{\bar uu+\bar dd}{\sqrt{2}} \rangle\sin \phi+|\bar ss\rangle \sin \phi,\qquad  \phi\simeq 39.3^0,
\end{eqnarray}
The corresponding expression for the form factors take the form 
\begin{eqnarray}
\label{Feta}
F_{\eta\gamma}(Q^2)=
\frac{N_c}{\sqrt2} \left(\frac49+\frac19\right)F_n(Q^2) \cos \phi-\frac{N_c}{9} F_s(Q^2)\sin\phi,\nonumber\\
F_{\eta'\gamma}(Q^2)=
\frac{N_c}{\sqrt2} \left(\frac49+\frac19 \right)F_n(Q^2)\sin \phi+\frac{N_c}{9}F_s(Q^2)\cos\phi. 
\end{eqnarray}
Here $F_n(Q^2)$ and $F_s(Q^2)$ are the form factors describing the transition of the nonstrange and 
$\bar ss$-components, respectively. 
The corresponding LD sum-rule for these form factors have a simple form 
\begin{eqnarray}
F_{n\gamma}(Q^2)&=&\frac{1}{f_n}\int\limits_0^{s_{\rm eff}^{(n)}(Q^2)}ds\,\Delta_{n}(s,Q^2),\nonumber\\
F_{s\gamma}(Q^2)&=&\frac{1}{f_s}\int\limits_0^{s_{\rm eff}^{(s)}(Q^2)}ds\,\Delta_{s}(s,Q^2).
\end{eqnarray}
$\Delta_n$ and $\Delta_s$ correspond to $\Delta_0$ with different quark masses in the loop. 
In numerical calculations we set $m_u=m_d=0$ and $m_s=100$ MeV.  
Accordingly, the LD model involves two separate effective
thresholds for the nonstrange and the strange components \cite{feldmann}:
\begin{eqnarray}
s_{\rm eff}^{(n)}&=4\pi^2f_n^2,\qquad
f_n\approx1.07f_\pi,\qquad 
s_{\rm eff}^{(s)}&=4\pi^2f_s^2,\qquad f_s\approx1.36f_\pi.
\end{eqnarray}
The LD model may not perform well for small values of $Q^2,$ where the
true effective threshold is smaller than the~LD threshold.
However, for larger $Q^2$ the LD model gives reasonable predictions
for the form factors, as illustrated by~Fig.~\ref{Plot:3}.

\vspace{-.3cm}

\section{Conclusions}
In this paper, we analyzed the $P\to\gamma\,\gamma^*$ transitions for $P=\pi^0,\eta,\eta'$. 

\noindent I. We emphasized that the $P\to\gamma\,\gamma^*$ form factor is proportional to the axial anomaly only if 
both photons are on-shell; if at least one of these photons is virtual, this proportionality is lost. 
As a result, the $P\to\gamma\,\gamma^*$ form~factor~in~the chiral limit is not protected from receiving
higher-order radiative corrections by the Adler--Bardeen theorem. 
Moreover, for virtual photons, the one-loop expression for 
the $\langle VAV\rangle$ amplitude does not develop a pole at $p^2=0$, related to a massless pseudoscalar. 
Therefore, the one-loop result for the form~factor $F_0$ cannot represent the full result for this quantity.   
$F_0$ should receive radiative corrections at higher orders in the loop expansion, 
in spite of the absence of two-loop radiative corrections to the $\langle VAV\rangle$
amplitude reported in \cite{teryaev1,2loop}. 

Interestingly, the form factor $F_0$ and the axial anomaly are given by dispersive integrals involving the same 
function, $\Delta_0(s,q^2)$. By the argument given above, radiative corrections (coming from three and more loops)
to the spectral density $\Delta_0(s,q^2)$ cannot vanish. Then, the Adler--Bardeen theorem 
requires some specific conspiral properties of multiloop corrections to $\Delta_0(s,q^2)$ 
enforcing the vanishing of their integrals over $s$. 

\vspace{.2cm}
\noindent II. We applied the local-duality version of QCD sum rules to the $(\pi^0,\eta,\eta')\to\gamma\,\gamma^*$ 
transition form factors. An attractive feature of this approach is the possibility to study form factors of 
bound states without knowing subtle details of their structure. Moreover, it allows one to consider on equal footing 
form factors of different bound states. Our findings may be summarized as follows:

1.~We tested the accuracy of the LD model for the $P\gamma$ transition form factor in quantum mechanics. 
We calculated this form factor from the solution of the Schr\"odinger equation and compared with the result 
of the quantum-mechanical LD sum rule. This comparison reveals that for a usual bound state, with a
typical hadronic extension of about~1~fm,~the LD sum rule is expected to yield 
accurate predictions for the form factor for $Q^2$ larger
than a few GeV$^2$; this accuracy increases with $Q^2$. At small
but nonzero $Q^2$, deviations from the LD model depend on subtle
details of the confining interaction.

2.~Surprisingly, the {\sc BaBar} data for the pion form factor exhibit an extreme violation of local duality in the
$\pi\gamma$~form factor even at $Q^2=40$ GeV$^2$. Moreover, the violation of local duality increases with $Q^2$ in the
range~$Q^2=10-40$~GeV$^2.$

3.~Even more surprisingly --- taking into account the strong disagreement in the pion case --- the LD predictions  
agree with the experimental data for both the $\eta$ and $\eta'$ mesons in the rather broad range $Q^2=4-100$~GeV$^2.$

The question why the nonstrange component in $\eta$ and $\eta'$,
on the one hand, and the pion, on the other hand, should lead to a
qualitatively different behaviour of the $P\to\gamma\,\gamma^*$
form factor remains mysterious. So far no compelling
theoretical explanation for this strange phenomenon has been found (see also \cite{mikhailov,roberts}).

\vspace{4ex}{\it Acknowledgments.}
We are deeply indebted to Berthold Stech for his encouragement and active 
participation at initial stages of this work. 
We are grateful to I.~Balakireva, D.~Becirevic, S.~Mikhailov, 
O.~Nachtmann, and O.~Teryaev for valuable discussions.
D.~M.\ was supported by the Austrian Science Fund (FWF) under
Project No.~P22843 and is grateful to the Alexander von
Humboldt-Stiftung and the Institute of Theoretical Physics of the
Heidelberg University for financial support and hospitality during
his stay in Heidelberg, where this work was started.

\end{document}